# Large array of Schrödinger cat states facilitated by an optical waveguide

Wui Seng Leong[1,†], Mingjie Xin[1,†], Zilong Chen[1,†], Shijie Chai[1], Yu Wang[1] & Shau-Yu Lan[1,*]

[1]Division of Physics and Applied Physics, School of Physical and Mathematical Sciences, Nanyang Technological University, Singapore 637371, Singapore

* To whom correspondence should be addressed. Email: sylan@ntu.edu.sg
† These authors contributed equally to this work.

**Quantum engineering using photonic structures offer new capabilities for atom-photon interactions for quantum optics and atomic physics, which could eventually lead to integrated quantum devices. Despite the rapid progress in the variety of structures, coherent excitation of the motional states of atoms in a photonic waveguide using guided modes has yet to be demonstrated. Here, we use the waveguide mode of a hollow-core photonic crystal fibre to manipulate the mechanical Fock states of single atoms in a harmonic potential inside the fibre. We create a large array of Schrödinger cat states, a quintessential feature of quantum physics and a key element in quantum information processing and metrology, of approximately 15000 atoms along the fibre by entangling the electronic state with the coherent harmonic oscillator state of each individual atom. Our results provide a useful step for quantum information and simulation with a wide range of photonic waveguide systems.**

**Introduction**

The quantum engineering toolbox developed from free space atom-light interaction experiments over the decades has led to many exciting breakthroughs in the areas of quantum information, metrology, and simulation. Departing from free space interactions towards photonic system interfaces could provide new paradigms[1-17], such as increased interaction strength, scalability of the devices, and novel design of functionalities. For instance, atoms trapped by the evanescent fields of nanofibres and photonic crystal slabs have been used to study long-range atom-atom interactions mediated by the coupling of superradiant emission into the structure[4,14]. In quantum communication applications, single collective excitation[6] and light storage[9-11] have been demonstrated in nanofibres and hollow-core fibres. Moreover, tight and diffraction-free confinement of atoms in photonic structures was used for optical switching[7,17]. In metrology, cold atoms in hollow-core fibres have also shown potential applications in timekeeping and sensing[8,12,13]. Regardless of these achievements, quantum control and manipulation of both external and internal degrees of freedom of atoms using photonic waveguide modes has not been realised.

Here, we demonstrate coherent excitation of the Fock states of atoms in an optical harmonic potential formed by the fundamental $LP_{01}$ mode of a hollow-core photonic crystal fibre. We implement an anti-Jaynes-Cummings-type Hamiltonian to excite the coherent harmonic oscillator states[18] (denoted as a coherent state here) and create a Schrödinger cat (SC) state[18,19] using the $LP_{01}$ mode. Our realisation of the SC states is based on a 3 mm array of atoms trapped in an optical lattice potential inside a hollow-core fibre, as shown in Fig. 1. The diffraction-free optical waveguide allows us to prepare an array of harmonic potentials with nearly identical axial trapping frequency. The axial vibrational energy levels of the lattice form the Fock state basis |n> for our experiments.

**Results**

**Ground state cooling to the Zeeman insensitive state**. We first collect an ensemble of $^{85}$Rb atoms by Doppler and sub-Doppler cooling 5 mm above a 4-cm-long open-end hollow-core photonic crystal fibre. The fibre is a hypocycloid-shaped photonic crystal fibre with $1/e^2$ mode field radius of 22 μm[11-13]. Atoms are then transported into the fibre at a velocity of 2 cm s$^{-1}$ by an optical conveyer belt using a moving optical lattice formed by a pair of counterpropagating beams[13]. The optical lattice has a period of 410.5 nm, with a power of 110 mW per beam. When atoms are in the fibre, we hold the atoms in a stationary lattice and optically pump them into the magnetic-field-insensitive |F=2, m=0> state to avoid the influence of inhomogeneous magnetic fields, where F denotes the hyperfine ground state of $^{85}$Rb and m is the Zeeman state. The vibrational frequency of the harmonic-like trap formed by each lattice site is $\omega=2\pi\times400$ kHz in the axial direction and $\omega_r=2\pi\times3.5$ kHz in the radial direction.

To prepare atoms in the axial vibrational ground state of the optical harmonic potential, we implement Raman sideband cooling (RSC) to cool atoms to the |F=2, m=0, n=0> state[20]. The cooling cycle starts with exciting atoms from |F=2, m=0, n> to |F=3, m=0, n-1> with a pair of linearly and orthogonally polarised Raman lasers (RB1 and RB2) at 821 nm, as shown in Fig. 2a, where a magnetic field of 2 G is applied along the fibre axis to define the quantisation axis and break the Zeeman degeneracy. A linearly polarised depump beam is used to bring atoms back to the |F=2, m, n-1> state in the Lamb-Dicke regime, where the coupling between the internal spin states and motional states is strongly suppressed during spontaneous emission, to preserve the vibrational quantum number. Finally, a π-polarised optical pump beam is applied to accumulate atoms in the |F=2, m=0, n-1> state to complete the cooling cycle. The cooling process continues until all the atoms are in the |F=2, m=0, n=0> state, which is the dark state for all the laser beams. Figure 2b shows the vibrational spectroscopy before and after cooling. The mean vibrational quantum number <n> is determined by taking the ratio of the areas of the first red sideband $A_{rsb}$ and first blue sideband $A_{bsb}$ as <n>= $(A_{rsb}/A_{bsb})/(1-(A_{rsb}/A_{bsb}))$. We achieve <n>$_0$=0.25 after cooling from <n>=3.3 before cooling. An optical depth (OD) of one corresponds to approximately $1.5\times10^4$ atoms, which gives an average of 2 atoms per lattice site[12].

**Study of the coherence between Fock states**. To study the coherence of the spin-Fock states in our system, we use RB1 and RB2 detuned by 3 GHz−$\omega$/2$\pi$ to excite a two-photon transition between |F=2, m=0, n>≡|↓, n> and |F=3, m=0, n-1>≡|↑, n-1> through a $\pi$/2-$\pi$-$\pi$/2 spin-echo sequence. We observe 37% and 22% contrasts at 0.1 ms and 0.9 ms separation times between the $\pi$/2 and $\pi$ laser pulses, respectively, as shown in Fig. 3a. In our differential light shift compensated optical lattice potential[13], the spin coherence time of the two spin states |↑> and |↓> measured by a microwave $\pi$/2-$\pi$-$\pi$/2 spin-echo sequence exceeds tens of milliseconds. The decay of the contrast is, thus, mostly due to the amplitude noise of the optical lattice and the radial motion of atoms.

**Preparation of a quantum coherent harmonic oscillator state**. The oscillation amplitude of a classical harmonic oscillator increases when the oscillator is subject to an external sinusoidal force at its natural frequency. Similarly, for a quantum harmonic oscillator, resonant driving can be achieved when the oscillation frequency of the force is at the vibrational frequency of the quantum harmonic oscillator[21]. The oscillatory force in our experiment is created by the time-dependent dipole force from a pair of counterpropagating driving lasers whose relative frequency $\omega$' is tuned near the vibrational frequency $\omega$ of the oscillator. In the quantum mechanical formulation, this corresponds to an initial |n=0> state displaced by displacement operator $D(\alpha)$ to a coherent state

$|\alpha\rangle$. The coherent state created by applying the time-dependent dipole force for a time $t$ can be written as $\alpha=(\eta\Omega/\delta)\sin(\delta t/2)e^{-i\delta t/2}$, where $\eta=\sqrt{\omega_R/\omega}=0.20$ is the Lamb-Dicke parameter, $\omega_R$ is the recoil frequency of the driving lasers, $\Omega$ is the ac Stark shift of the atom internal state arising from the driving lasers, and $\delta=\omega-\omega'$ (see Methods).

The coherent state created by a harmonic oscillator can be written in the Fock state basis with a Poissonian distribution as $|\alpha\rangle=\exp(-|\alpha|^2/2)\Sigma_n \alpha^n/\sqrt{n!}|n\rangle$. Therefore, its amplitude can be determined by the mean phonon number $\langle n\rangle$ obtained from vibrational spectroscopy measurements. Figure 3b shows the amplitude of the coherent state $|\alpha|$ versus the laser driving time $t$, where atoms are initially prepared in $|\downarrow\rangle$ state. The pair of driving lasers have wavelengths of 795 nm and are 3.78 GHz red-detuned from the $|F=2\rangle$ to $|F'=2\rangle$ transition, with approximately 13 μW per beam. They are linearly polarised and relatively detuned by frequency $\omega'$. To avoid the effect of the radial motion while creating a coherent state, we operate the driving lasers in the short pulse regime $t\ll 1/\omega_r$. When $\delta t$ is small, the amplitude of the coherent state is linear with the driving time $t$ and independent of the detuning $\delta$ as $\alpha=\eta\Omega t/2$. The data show a linear trend confirming the approximation of our system.

However, this amplitude measurement does not reveal information about the coherence between the superposition of Fock states in the coherent state[22]. We demonstrate the coherence in the coherent state by applying two consecutive driving lasers pulses with phase difference $\phi$. The mean phonon number after the application of two successive displacement operators $\hat{D}(\alpha)$ and $\hat{D}(\alpha e^{i\phi})$ is $\langle n\rangle=\langle n\rangle_0+2|\alpha|^2+2|\alpha|^2\cos\phi$, as shown in Fig. 3c. The mean phonon number returning to the initial value $\langle n\rangle_0$ when the phase $\phi$ is $\pi$ indicates the coherence between different Fock states in the coherent state. The fitted data show the coherent state with an amplitude $|\alpha|=0.36$.

**Creation of an array of the SC states**. The SC state is a superposition of two spatially separated but localised classical states entangled with an auxiliary superposition state, where the projection measurement on the auxiliary state (or atomic decay) collapses the SC state into one or the other widely separated classical states (or live or dead cat)[19]. The SC state can be created by entangling a superposition of coherent states with the internal spin states of an atom. For trapped ions SC states, the coherent state α has been created in ions' motional Fock states basis[18,23-27] as large as |α|=12. In SC states with neutral atoms[28-31] and superconducting circuits[32], the coherent states have been formed by the photonic Fock states in resonators with |α| around 1-3. These approaches, typically including one SC state in single apparatus, have been used for quantum information processing and for testing the boundary between classical and quantum physics[18,28]. For neutral atoms' motional Fock states, the coherent states have been demonstrated in a harmonic potential[33,34]. Here, we create the SC states in an optical lattice by entangling the coherent states and spin states in an optical waveguide.

The protocol of the SC state preparation is shown in Fig. 4a[23]. A microwave $\pi/2$ pulse of 7.5 μs brings the atoms from the vibrational ground state |↓, *n*=0> into a superposition state of spin up |↑, *n*=0> and spin down |↓, *n*=0>. The coherent state |↑, α> is prepared by the driving lasers, whose beat frequency is resonant on the vibrational frequency. The internal spin states are then flipped by a microwave $\pi$ pulse, followed by another driving lasers pulse with a different phase $\phi$ to create a coherent state. The resulting SC state can then be written as

$$\frac{-|\downarrow\rangle|\alpha\rangle - i|\uparrow\rangle|\alpha e^{i\phi}\rangle}{\sqrt{2}}. \qquad (1)$$

Equation (1) is an entangled state between a superposition of coherent states and spin states. To confirm this, we measure the interference of the two separated coherent states by applying another microwave π/2 pulse with a phase shift $\delta_M$ relative to the first two microwave pulses, and the state becomes

$$i|\uparrow\rangle\left(\frac{e^{i\delta_M}|\alpha\rangle-|\alpha e^{i\phi}\rangle}{2}\right)-|\downarrow\rangle\left(\frac{|\alpha\rangle+e^{-i\delta_M}|\alpha e^{i\phi}\rangle}{2}\right). \quad (2)$$

The population of atoms in state |↑⟩ is then determined by the overlap between the two coherent states in phase space. We detect the population of the atoms in state |↑⟩ by absorption detection (OD) with varying phase $\phi$. The probability of finding the atoms in |↑⟩ is

$$P_\uparrow=\frac{1}{2}\left(1-e^{-|\alpha|^2(1-\cos\phi)}\cos(\delta_M+|\alpha|^2\sin\phi)\right). \quad (3)$$

Figure 4b shows the experimental data of SC state interference fringes for different driving lasers pulse durations. The population of the atoms in |↑⟩ is normalised to 0.5 when the driving lasers are off, and the relative phase is set at $\delta_M=\pi/2$. When the microwave phase is set to $\delta_M=0$ ($\pi$), the SC state is termed the odd (even) cat state. The odd (even) cat states only contain odd (even) Fock states, similarly to squeezed vacuum states. We use this approach to verify the validity of the SC state created in our experiment. Figure 4c shows a comparison of the SC states when $\delta_M=0$ and $\pi$, and it clearly confirms the trend of population change of Eq. (3) for odd (even) cat states.

Our implementation of the generation of SC states deviates from the ideal case of Eq. (3) due to the imperfect conditions of our experiment. We modify Eq. (3) to fit the data in Fig. 4b by introducing the effects of the residual excitation of the driving lasers on the |↓⟩ state, contrast $C$, and imperfect initial ground state preparation as

$$P_\uparrow=\frac{1}{2}\left(1-Ce^{-(2\langle n\rangle_0+1)|\beta|^2(1-\cos\theta)}\cos(\delta_M+|\beta|^2\sin\theta)\right), \quad (4)$$

where $\beta=(1+\varepsilon e^{i\phi'})\alpha$, $e^{i\theta}=(e^{i\phi'}+\varepsilon)/(1+\varepsilon e^{i\phi'})$, $\phi'=\phi-\phi_f$, and $\phi_f=\omega'\Delta t$ is the phase advance of the driving lasers due to the increase in the pulse time $\Delta t$. For the residual driving, the parameter $\varepsilon=0.19$ is used to consider the off-resonant dipole force on $|\downarrow\rangle$ (see Methods). The imperfect ground state preparation is taken into account by adding a factor of $2\langle n\rangle_0+1=1.5$ into the exponent of Eq. (3). The fitting parameters $|\alpha|$, $C$, and $\phi_f$ are displayed in Fig. 4d. The phase advance $\phi_f$ is linear in time, as expected, due to the free evolution of the phase of the beating of the driving lasers. For $t=0.45$ μs, the amplitude of the coherent state is $\alpha=1.42$, which corresponds to a maximum separation of the two coherent wave packets of $2\sqrt{2}|\alpha|z_0=54$ nm, where $z_0^2=(\hbar/2m\omega)$ is the variance in the ground state wave packet size, $\hbar$ is the reduced Planck constant, and $m$ is the mass of $^{85}$Rb. The fidelity of the state can be estimated from the contrast[27] as $C^{1/2}$. The loss of contrast suggests phase noise during the operation[23], which also widen the narrow interference feature. The increase of the interference width leads to the saturation of $|\alpha|$ after $t=0.45$ μs for the fitting of Eq. (4).

We study the effect of radial motion during the free evolution between driving lasers pulses by inserting an additional free evolution time $\tau$, as shown in Fig. 4e. The sequence follows $\pi/2$-$\hat{D}(\alpha)$-$\tau$-$\pi$-$\hat{D}(\alpha e^{i\phi})$-$\pi/2$. The data show that the radial motion does not significantly degrade the contrast at our experimental time scale. We also numerically integrate Eq. (4) over the radial distribution of atoms with radial position-dependent parameters of $\eta$, $\Omega$, and $\delta$, as shown in Fig. 4f (see Methods). Our results indicate that the radial distribution of atoms is not the main loss mechanism for the interference contrast. The anharmonicity of the optical lattice also causes deviation of the vibrational energy splitting between $|n\rangle$ and $|n+1\rangle$ from the pure harmonic potential as $\omega_{anh}=\omega-\omega_{rec}(1+n)$, where $\omega_{rec}$ is the recoil frequency of the lattice. This gives an approximately few percent deviation at the centre of the fibre[35]. Other decoherence mechanisms, such as single-photon scattering of the driving lasers and amplitude fluctuation of the lattice potential[36], are 10 and 100 times slower than the coherent interaction between the driving lasers and the atoms.

**Discussion**

Our results show the coherent interaction of single quantum harmonic oscillator states of matter with the light guided in a fibre. We use the fundamental waveguide mode to create entanglement between coherent states and spin states inside a hollow-core fibre. Such a state can enable quantum simulation of a truly one-dimensional system with the assistance of ground state cooling in the radial direction[1,37]. Other than the fundamental importance, the plethora of photonic structures of fibres[38] provides fruitful ways for manipulating atoms. For example, the higher-order modes can be tailored to create multiple lattice sites and different polarisations in the transverse plane, creating a three-dimensional atom array in a cylindrically symmetric photonic structure[39]. Strong atom-light interactions can be achieved by trapping atoms in the cladding modes, which can be patterned by engineering the geometry and thickness of the cladding structure[40]. As silica fibreglass is transparent to light in the visible to near-infrared range, light propagating from the side into the fibre can also be used for controlling the internal and external degrees of freedom of atoms[41]. Other photonic crystal structures also possess similar degrees of freedom to control atoms at the single quantum state level. Our experiments pave the way for the realisation of quantum devices of photonic structures with matter for promising applications in quantum information science.

**Methods**

**Experimental details**. Cold atoms in this experiment are prepared by Doppler and sub-Doppler cooling of a magneto-optical trap (MOT) 5 mm above the fibre tip in 1 s. At this stage, the MOT contains approximately $10^8$ atoms at 8 μK. After switching off the cooling lasers, a moving optical lattice is turned on to transport atoms into the fibre at a velocity of 2 cm s$^{-1}$, where the velocity is determined by the frequency difference between the two counterpropagating laser beams. When atoms are a few mm inside the fibre after 200 ms of transport, the frequency difference of the lasers is ramped down to zero to form a stationary lattice. The lattice contains approximately $2\times10^4$ atoms. We subsequently apply a push beam in a cycling transition to blow away any atoms above the fibre tip to ensure that only the atoms inside the fibre will be probed. The population in the $F=3$, $m=0$ state is measured by the transmission $T$ of a 50 μs probe pulse in the $F=3$ to $F'=4$ $D2$ transition and represented by the optical depth (OD) as OD=−ln($T$).

The two optical lattice beams are from an 821 nm Ti-sapphire laser. They separately pass through two independent acoustic-optical modulators (AOMs) to control their relative frequencies. RB1 is formed by sending part of the Ti-sapphire laser to a 3 GHz electro-optical modulator, followed by a temperature-stabilised solid-state etalon that selects the +1$^{st}$ order laser light. It then passes through an AOM for switching and frequency shifting. The driving lasers are from an extended cavity diode laser frequency locked at 0.744 GHz from $F=3$ to $F'=2$ on the $D1$ line. It is split into two paths with two independent AOMs before being coupled into the fibre from both ends.

**Derivation of the displacement operator**. The Hamiltonian of a forced harmonic oscillator in the interaction picture has a form similar to the classical harmonic oscillator

$$\hat{H}(t)=\hbar(f^*(t)z_0\hat{a}e^{-i\omega t}+f(t)z_0\hat{a}^\dagger e^{i\omega t}), \qquad (5)$$

where $f(t)$ is a time-dependent force and $\hat{a}$ and $\hat{a}^\dagger$ are the annihilation and creation operators, respectively. The state after some interaction time $t$ can be characterised by the time-evolution operator

$$\hat{U}(t)=\exp\left(-\frac{i}{\hbar}\int_0^t \hat{H}(t')\,dt'\right)\equiv \hat{D}(\alpha)=e^{\alpha\hat{a}^\dagger+\alpha^*\hat{a}}, \qquad (6)$$

where we only consider the first term in the exponent and $\alpha$ is defined as

$$\alpha(t)=-\frac{i}{\hbar}\int_0^t f(t)z_0 e^{i\omega t'}\,dt'. \qquad (7)$$

The time-dependent force $f(t)$ in our experiment is provided by a moving standing wave resulting from a pair of counterpropagating lasers whose frequency difference $\omega'$ is set to be near the vibrational frequency $\omega$ of the harmonic potential. The interaction Hamiltonian between the spin states $|\uparrow\rangle$ of atoms and the lasers can be written as

$$\hat{H}(t)=\frac{\hbar}{2}\left[\Omega e^{i(\mathbf{k}_{\text{eff}}\cdot\mathbf{z}-\omega' t-\phi)}+h.c.\right]|\uparrow\rangle\langle\uparrow|, \qquad (8)$$

where $\Omega$ is the Rabi frequency of the driving lasers, $k_{\text{eff}}$ is the effective wavenumber of the lasers, $\omega'$ is the relative frequency of the driving lasers, and $\phi$ is the relative phase between the two lasers. In the Lamb-Dicke regime, where $\mathbf{k}_{\text{eff}}\cdot\mathbf{z}= k_{\text{eff}}z_0(\hat{a}^\dagger e^{i\omega t}+ \hat{a}e^{-i\omega t}) \ll 1$, $\exp[i\mathbf{k}_{\text{eff}}\cdot\mathbf{z}]\approx(1+i\eta(a^\dagger e^{i\omega t}+ ae^{-i\omega t})-O(\eta^2))$, where $\eta=k_{\text{eff}}z_0$. By detuning the driving lasers from the vibrational frequency, the corresponding Hamiltonian for the first blue sideband is

$$\hat{H}(t)=\frac{\hbar}{2}\Omega\eta[\hat{a}^{\dagger}e^{i(\delta t-\phi)}+\hat{a}e^{-i(\delta t-\phi)}]|\uparrow\rangle\langle\uparrow|, \tag{9}$$

where $\delta=\omega-\omega'$.

The anti-Jaynes-Cummings-type Hamiltonian has the same form as a forced quantum harmonic oscillator, and $\alpha(t)$ can now be written as

$$\alpha(t)=\frac{\eta\Omega}{\delta}\sin(\frac{\delta}{2}t)e^{-i\frac{\delta}{2}t}e^{i\phi}. \tag{10}$$

**Schrödinger cat state interferometer**. In the preparation of the coherent harmonic state, ideally, the driving lasers should only excite atoms in $|\uparrow, n=0\rangle$. However, off-resonant excitation of $|\downarrow, n=0\rangle$ also occurs. We model this residual excitation by the coefficient $\varepsilon=\Delta_s/(\Delta_s+f_{hf})$, where $\Delta_s=$ 744 MHz is the single-photon detuning of the driving lasers from the $|F=3\rangle$ to $|F'=2\rangle$ D1 transition, and $f_{hf}=3$ GHz is the frequency splitting of $|F=3\rangle$ and $|F=2\rangle$. After the first microwave $\pi/2$ pulse, the driving lasers pulse operates on $|\downarrow, n=0\rangle$ and $|\uparrow, n=0\rangle$ with $\varepsilon\alpha$ and $\alpha$, respectively, and the state becomes

$$\sqrt{\frac{1}{2}}|\downarrow, \varepsilon\alpha\rangle - i\sqrt{\frac{1}{2}}|\uparrow, \alpha\rangle. \tag{11}$$

A $\pi$ microwave pulse flips the spin, and the second driving lasers pulse operates on $|\downarrow, \alpha\rangle$ and $|\uparrow, \varepsilon\alpha\rangle$ with $\varepsilon\alpha e^{i\phi'}$ and $\alpha e^{i\phi'}$, respectively, where $\phi'=\phi+\omega'\Delta t$, $\phi$ is the relative phase of the driving lasers and $\omega'\Delta t$ is the phase advance during the extra waiting time between driving lasers pulses. The state then becomes

$$e^{i\varepsilon|\alpha|^2\sin\phi'}[-i\sqrt{\frac{1}{2}}|\uparrow, (e^{i\phi'}+\varepsilon)\alpha\rangle - \sqrt{\frac{1}{2}}|\downarrow, (1+\varepsilon e^{i\phi'})\alpha\rangle] \tag{12}$$

The global phase term $\varepsilon|\alpha|^2\sin\phi$ can be ignored in our measurements. The final $\pi/2$ microwave pulse with phase $\delta_M$ relative to the first two microwave pulses creates the state

$$\frac{i}{2}(e^{i\delta_M}|\uparrow, (1+\varepsilon e^{i\phi'})\alpha\rangle-|\uparrow, (e^{i\phi'}+\varepsilon)\alpha\rangle)-\frac{1}{2}(|\downarrow, (1+\varepsilon e^{i\phi'})\alpha\rangle+e^{-i\delta_M}|\downarrow, (e^{i\phi'}+\varepsilon)\alpha\rangle) \quad (13)$$

$$\equiv \frac{i}{2}(e^{i\delta_M}|\uparrow, \beta\rangle-|\uparrow, \beta e^{i\theta}\rangle)-\frac{1}{2}(|\downarrow, \beta\rangle+e^{-i\delta_M}|\uparrow, \beta e^{i\theta}\rangle),$$

where we define $\beta\equiv(1+\varepsilon e^{i\phi'})\alpha$ and $e^{i\theta}\equiv(e^{i\phi'}+\varepsilon)/(1+\varepsilon e^{i\phi'})$. The population of atoms in the $|\uparrow\rangle$ state $P_\uparrow$ is

$$P_\uparrow = \frac{1}{2} - \frac{1}{2}e^{-|\beta|^2(1-\cos\theta)}\cos(\delta_M+|\beta|^2\sin\theta). \quad (14)$$

**Temperature effect in the Schrödinger cat state interferometer**. Due to the finite temperature of the atoms in the radial direction, atoms at different locations of the radial trap experience different axial trapping frequencies and therefore different Lamb-Dicke parameters and Rabi frequencies of the driving lasers. Assuming that the above parameters have a Gaussian distribution radially, we numerically integrate Eq. (3) over the position $r$-dependent coherent harmonic state $\alpha(r)=(\eta(r)\Omega(r)/\delta(r))\sin(\delta(r)t/2)e^{-i\delta(r)t/2}$ weighted by the Gaussian distribution of the atoms as

$$\frac{\int_0^\infty 2\pi r e^{\frac{-r^2}{w_a^2}}\left(\frac{1}{2}\left(1-Ce^{-(2\langle n\rangle_0+1)|\beta(r)|^2\times(1-\cos\theta)}\cos(\delta+|\beta(r)|^2\sin\theta)\right)\right)dr}{\int_0^\infty 2\pi r e^{\frac{-r^2}{w_a^2}}dr}, \quad (15)$$

where the centre of the fibre is defined as $r=0$, $w_a^2=W^2 k_B T_r/2/U$ is the $1/e$ radius of the spatial distribution of the atomic cloud in the radial direction, $W=22$ μm is the $1/e^2$ mode field radius, $k_B$ is the Boltzmann constant, $T_r$ is the radial atom temperature, and $U$ is the radial trapping potential. The results for different temperatures are shown in Fig. 4f.

## Acknowledgements

We thank Dzmitry Matsukevich for fruitful discussion and Thomas Billotte on the fibre modal characterisation. This work is supported by Singapore National Research Foundation under Grant No. NRFF2013-12 and Quantum Engineering Programme under Grant No. QEP-P4, and Singapore Ministry of Education under Grant No. MOE2017-T2-2-066.

## Author contributions

W. S. L., M. X., Z. C., S. C., Y. W., and S.-Y. L., contributed to designing the experiment, taking and analysing the data, and preparing the manuscript.

## Competing interests

The authors declare no competing interests.

## Data availability

The data that support the plots within this paper and other findings of this study are available from the corresponding author upon reasonable request.

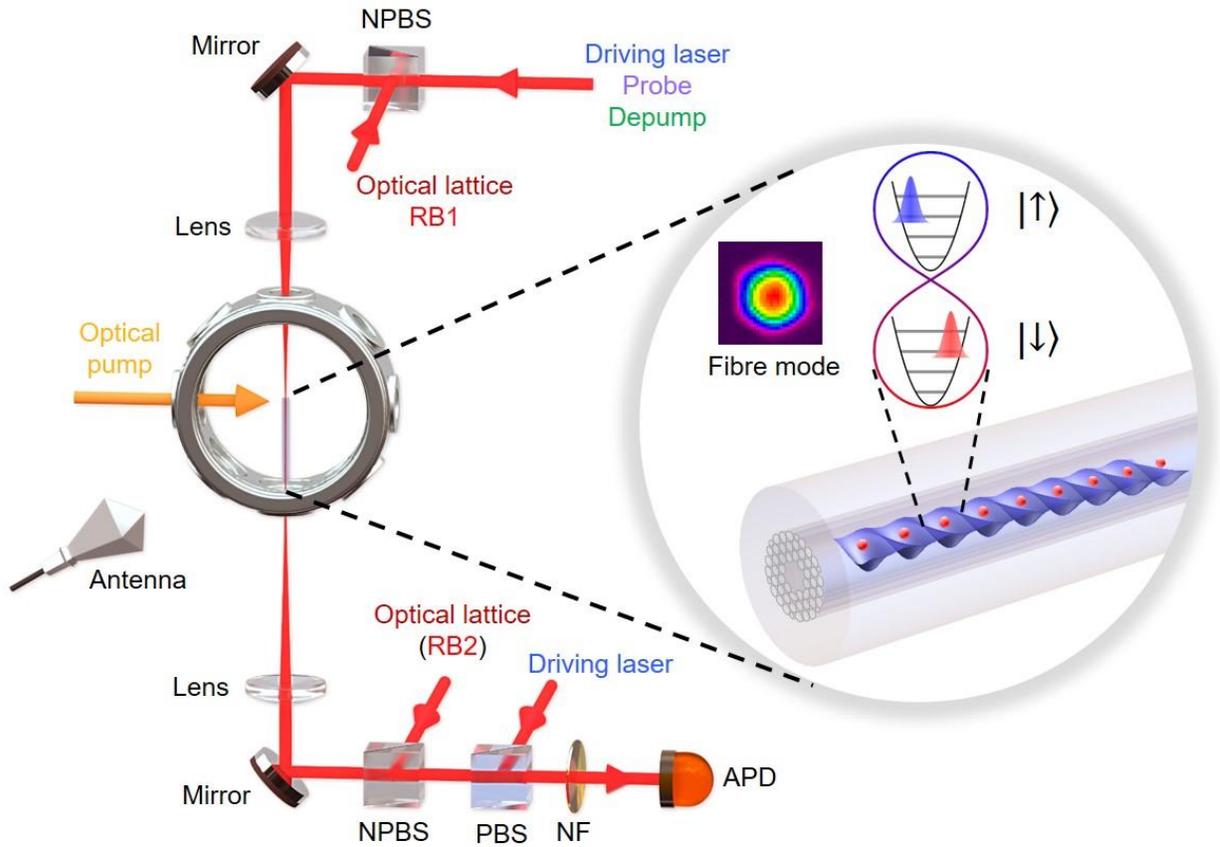

**Figure 1 | Schematic of the experimental setup.** Two 150 mm achromatic lenses are used to couple all the beams into the fibre with 70% coupling efficiency. The $\pi$-polarised optical pump beam incident from the side of the fibre can efficiently prepare atoms in the $F=2$, $m=0$ state with more than 90% efficiency. Each lattice site in the fibre represents a harmonic oscillator potential for atoms (red filled circles). The coloured Gaussian wave packets in the potential illustrate the coherent harmonic oscillator states that are entangled with the spin states. The inset shows a near-field image of the waveguide mode. PBS: polarising beam splitter. NPBS: nonpolarising beam splitter. NF: notch filter. APD: avalanche photodiode.

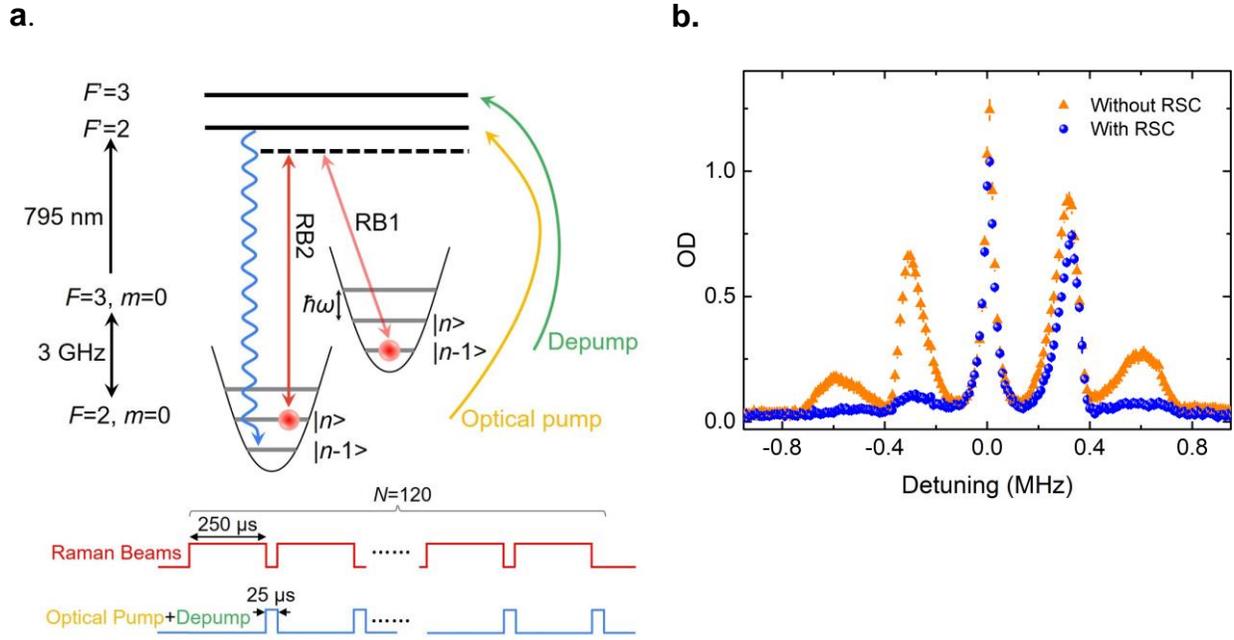

**Figure 2 | Raman sideband cooling in the fibre. a,** Energy levels and timing sequence. The energy levels represent the vibrational energy splitting of the optical lattice potential in the axial direction. The wavy blue arrow denotes the spontaneous emission of atoms. Each cooling cycle consists of 250 μs of the Raman pulses (RB1 and RB2) and 25 μs of the depump and optical pump pulses. The overall duration for the best cooling performance is $N$=120 cooling cycles. RB1 and RB2 are detuned by 3 GHz−$\omega/2\pi$ to reduce $n$ by 1. **b,** Vibrational spectra. We use RB1 and RB2 to excite atoms to the $F$=3, $m$=0 state. The $x$-axis is the detuning between the two Raman beams and is set at zero at the carrier transition frequency. We plot the OD for the $F$=3, $m$=0 state as a function of the detuning with and without RSC. Most of the atoms are accumulated in the $n$=0 state after RSC; therefore, the blue sidebands remain (mostly $n$=0 to $n$=1 transition), and the red sidebands are suppressed. Smaller OD after RSC is mainly due atom loss during RSC. Error bars are the standard error of mean of 4 experimental runs.

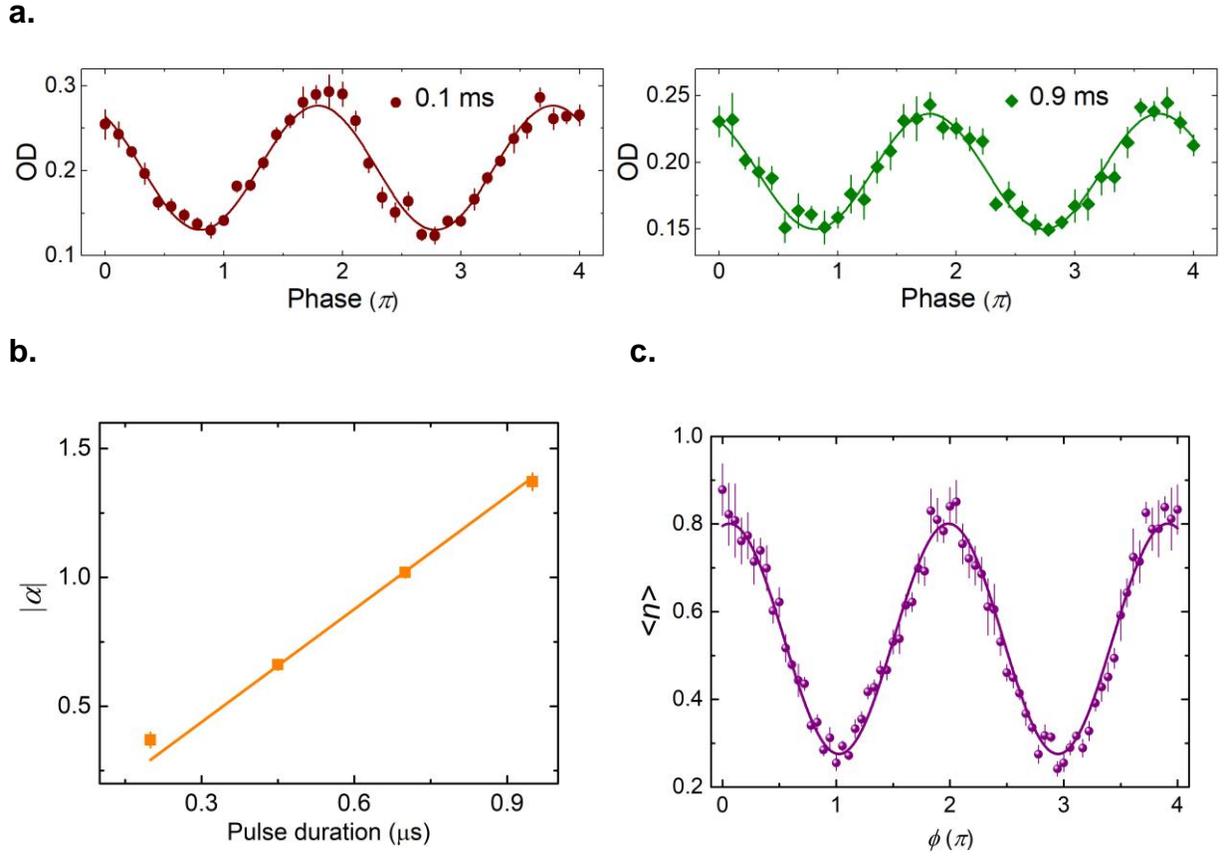

**Figure 3 | Preparation of a quantum coherent harmonic oscillator state. a,** Coherence time measurements between $|\downarrow, n\rangle$ and $|\uparrow, n\text{-}1\rangle$ states. A $\pi/2$-$\pi$-$\pi/2$ spin-echo sequence is implemented on the atoms before sideband cooling. The duration of the $\pi/2$ laser pulses is 7.5 μs. The OD of $|\uparrow, n\text{-}1\rangle$ is measured with varying phase of the final $\pi/2$ laser pulse. Error bars are the standard error of mean of 4 experimental runs. The curves are sinusoidal fits to the data. **b,** Measurements of the amplitude of the coherent state as a function of driving lasers pulse duration, where atoms are initially prepared in $|\downarrow\rangle$ state. We determine the amplitude of the coherent state by measuring $<n>$ with vibrational spectroscopy, similar to Fig. 2b. Each data point corresponds to one vibrational spectroscopy measurement. Error bars are the standard error of mean of 4 experimental runs. The straight line is a linear fit to the data with a fixed intercept at zero. **c,** Measurements of $<n>$ of a coherent state after two consecutive driving lasers pulses with phase difference $\phi$. The coherence

of Fock states components in the coherent state is examined by applying two consecutive driving lasers pulses with phase difference $\phi$. After the two driving lasers pulses, we measure <$n$> from the vibrational spectra and plot it for different phases between the first and second displacement pulses. Error bars are the standard error of mean of 4 experimental runs. The curve is a sinusoidal fit to the data.

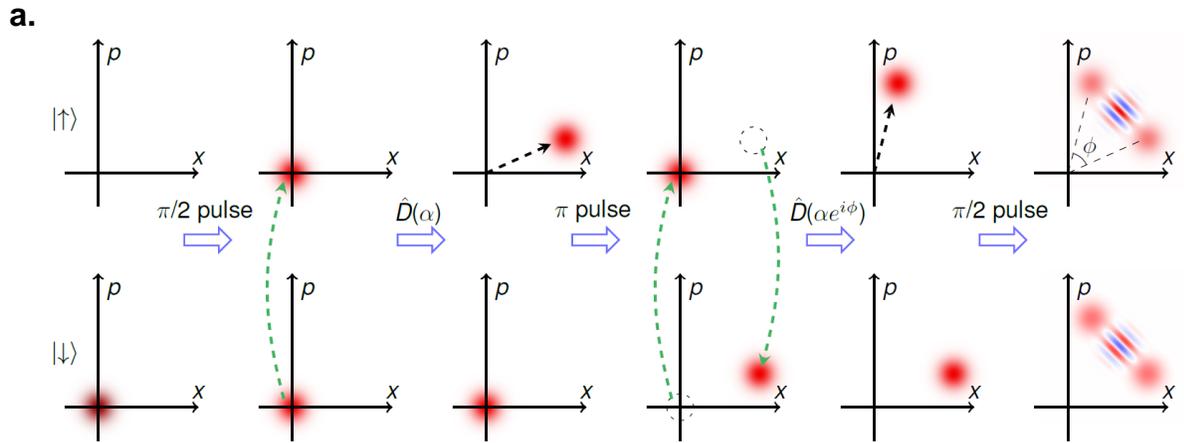

**a.**

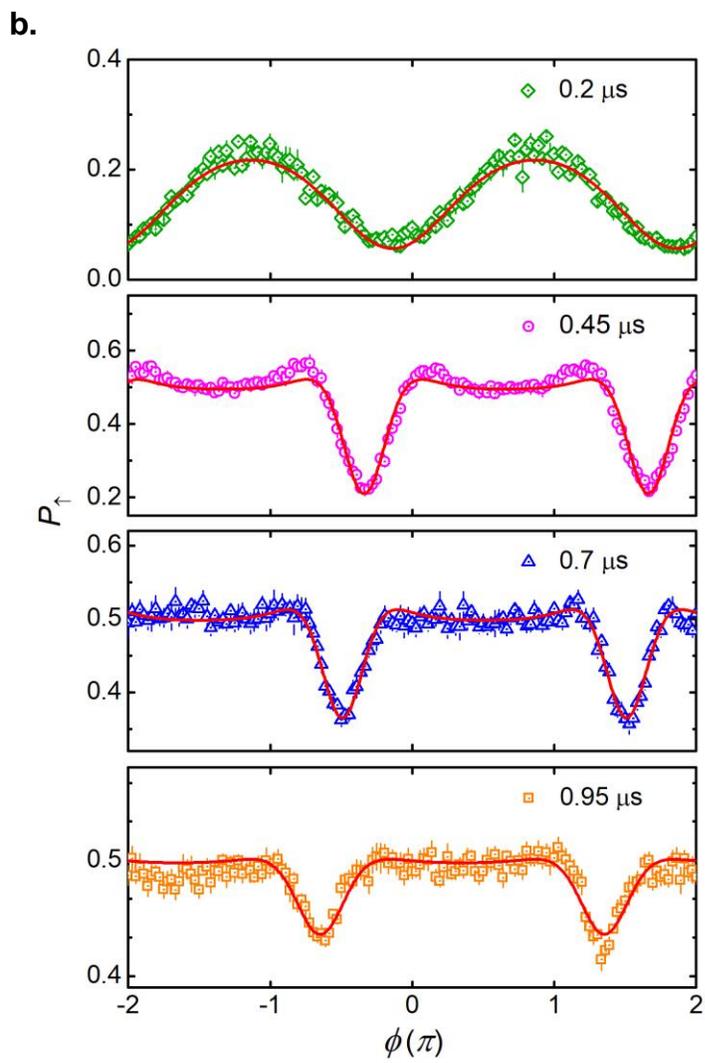

**b.**

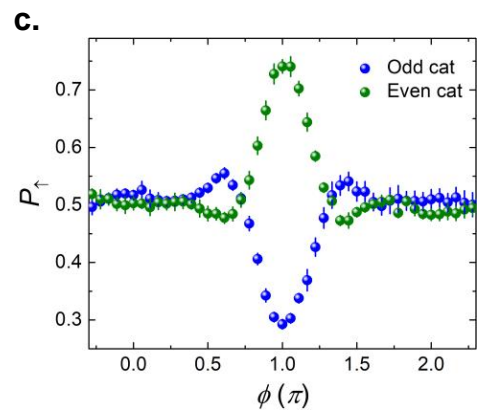

**c.**

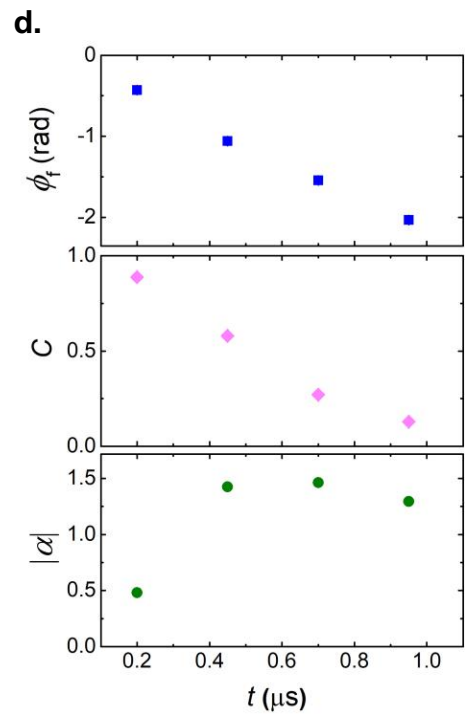

**d.**

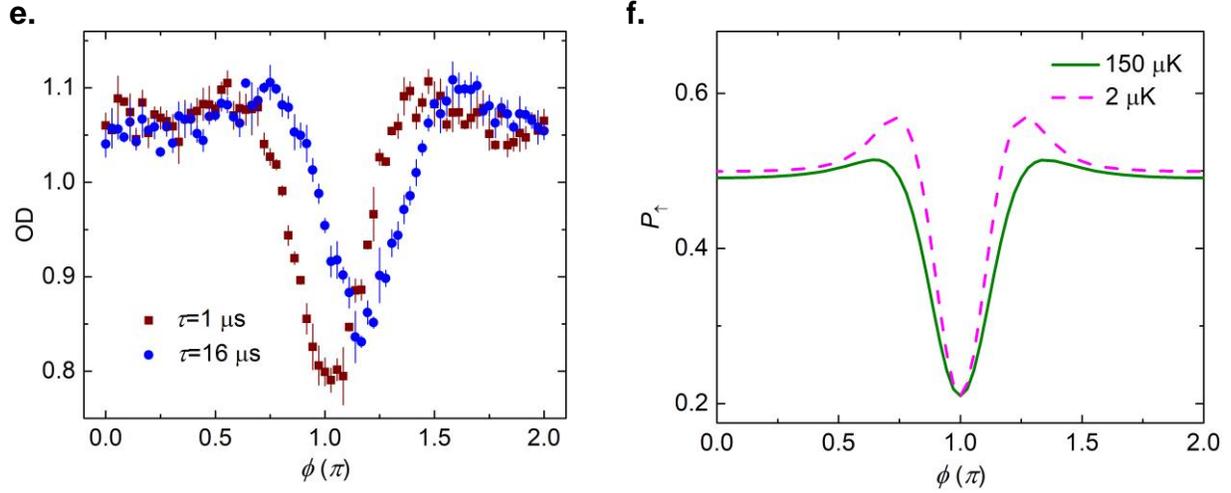

**Figure 4 | SC state interferometer. a,** Protocol for creating the SC state in phase space representation. The top and bottom columns represent |↑> and |↓>, respectively. The circles in red indicate the uncertainty in the ground state and the coherent state. The microwave π/2 and π pulses are used to manipulate the internal spin states. The displacement operator $\hat{D}(\alpha)$ and $\hat{D}(\alpha e^{i\phi})$ are created by driving lasers pulses. **b,** Interference fringes of the SC state interferometer. The population of atoms in |↑> state $P_\uparrow$ is measured with the relative phase $\phi$ of the two driving lasers pulses for different pulse durations. Error bars are the standard error of mean of 4 experimental runs. **c,** Even and odd SC states. The population of atoms in |↑> state $P_\uparrow$ is measured with the relative phase $\phi$ of the two driving lasers pulses for $\delta_M=0$ (odd SC state) and $\delta_M=\pi$ (even SC state). The driving lasers pulse duration is 0.45 μs. Error bars are the standard error of mean of 4 experimental runs. **d,** Fitting parameters of Eq. (4) for the experimental data in Fig. 4b is plotted against different driving lasers pulse duration. **e**. Measurements of the |↑> state population $P_\uparrow$ in cat interferometer operation versus phase $\phi$ with additional free evolution time $\tau$. The total free evolution time between the two driving lasers pulses for $\tau=1$ μs is 16 μs. Error bars are the standard error of mean of 4 experimental runs. **f**. Simulation of the temperature effect on the SC state interferometer. The simulation uses the fitted parameters for $t=0.45$ μs in Fig. 4d. The 1/$e$ Gaussian

radius of the atomic cloud in the radial direction is 0.9 μm and 8 μm for 2 μK and 150 μK, respectively.